\documentclass[conference]{IEEEtran}
\IEEEoverridecommandlockouts
\pdfoutput=1

\usepackage{cite}
\usepackage{amsmath,amssymb,amsfonts}
\usepackage{algorithmic}
\usepackage{graphicx}
\usepackage{textcomp}
\usepackage{booktabs}
\usepackage{url}
\usepackage{xcolor}
\def\BibTeX{{\rm B\kern-.05em{\sc i\kern-.025em b}\kern-.08em
    T\kern-.1667em\lower.7ex\hbox{E}\kern-.125emX}}
\begin{document}

\title{DiffCSS: Diverse and Expressive Conversational Speech Synthesis with Diffusion Models\\

\thanks{This work is supported by National Natural Science Foundation of China (62076144) and Shenzhen Science and Technology Program (WDZC20220816140515001, JCYJ20220818101014030).}
}

\author{

\IEEEauthorblockN{
{\begin{tabular}{c}
        \textit{Weihao Wu$^{1,3,*}$, Zhiwei Lin$^{1,*}$\thanks{*These authors contributed equally to this work as first authors.}, Yixuan Zhou$^{1}$, Jingbei Li$^{4}$, Rui Niu$^{1,3}$, Qinghua Wu$^{3}$}, \\ 
        \textit{Songjun Cao$^{3}$, Long Ma$^{3}$, Zhiyong Wu$^{1,2,\dagger}$
        \thanks{$\dagger$ Corresponding Author.}}
    \end{tabular}}}
\IEEEauthorblockA{$^1$ Shenzhen International Graduate School, Tsinghua University, Shenzhen, China\\
    $^2$ The Chinese University of Hong Kong, Hong Kong SAR, China\\
  $^3$ Tencent Youtu Lab \quad $^4$ StepFun \\
\{wuwh23, lzw22\}$@$mails.tsinghua.edu.cn, zywu@sz.tsinghua.edu.cn}
}

\maketitle

\begin{abstract}

Conversational speech synthesis (CSS) aims to synthesize both contextually appropriate and expressive speech, and considerable efforts have been made to enhance the understanding of conversational context.
However, existing CSS systems are limited to deterministic prediction, overlooking the diversity of potential responses.
Moreover, they rarely employ language model (LM)-based TTS backbones, limiting the naturalness and quality of synthesized speech.
To address these issues, in this paper, we propose DiffCSS, an innovative CSS framework that leverages diffusion models and an LM-based TTS backbone to generate diverse, expressive, and contextually coherent speech.
A diffusion-based context-aware prosody predictor is proposed to sample diverse prosody embeddings conditioned on multimodal conversational context.
Then a prosody-controllable LM-based TTS backbone is developed to synthesize high-quality speech with sampled prosody embeddings.
Experimental results demonstrate that the synthesized speech from DiffCSS is more diverse, contextually coherent, and expressive than existing CSS systems\footnote{Audio samples: \url{https://w-w-h.github.io/DiffCSS/}}.

\end{abstract}

\begin{IEEEkeywords}
Conversational speech synthesis, diffusion models, prosody diversity
\end{IEEEkeywords}

\section{Introduction}

With the development of deep learning, end-to-end text-to-speech (TTS) systems have made significant strides \cite{wang2017tacotron, shen2018natural, ren2020fastspeech, kim2021conditional}. However, in certain application scenarios such as chatbots and virtual assistants, these systems often underperform due to their limited ability to understand conversational context, highlighting the growing importance of conversational speech synthesis (CSS). CSS aims to generate speech that is not only appropriate for the current utterance but also coherent with the broader conversational context.
Therefore, effective context modeling is crucial for CSS models.  
Guo et al. \cite{guo2021conversational} are the first to propose a GRU-based conversational context encoder that sequentially processes the textual conversational context. 
Subsequently, graph-based networks \cite{li2022enhancing, li2022inferring} are introduced to model the cross-sentence and cross-speaker relationships. 
To enhance the extraction of fine-grained details, multi-scale information has also been explored \cite{li2022inferring, xue2023m}. Deng et al. \cite{deng2024concss} introduce contrastive learning to CSS, facilitating the learning of more stable and discriminative context representation. Most recently, Liu et al. \cite{liu2024generative} employ a GPT-based architecture to capture semantic and stylistic features within conversational sequences.
\begin{figure*}
  \hspace*{\fill}
  \includegraphics[scale=0.4]{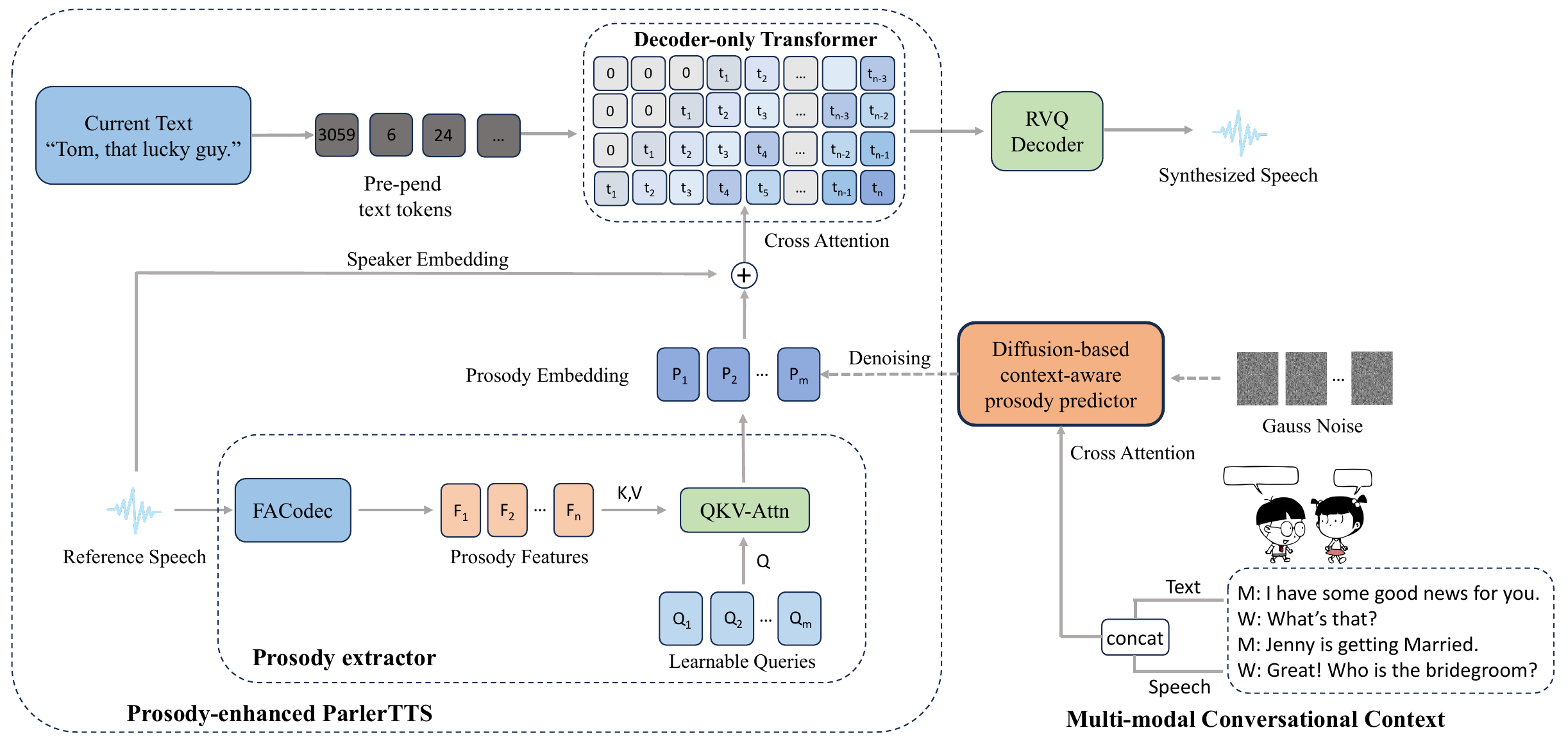}
  \hspace*{\fill}
  \vspace{-5pt}
  \caption{Overall architecture of the proposed CSS framework}
  \vspace{-10pt}
  \label{fig:overview}
\end{figure*}

However, current CSS methods are constrained by deterministic prosody predictions. 
Specifically, in a given conversational context, speech can be delivered in various ways, each conveying different emotions and intentions.
Consequently, multiple prosody variations can correspond to the same conversational context.
This one-to-many mapping problem has not been thoroughly explored in previous CSS systems, thus limiting their ability to generate diverse and expressive prosody. 

Additionally, nearly all previous CSS approaches have relied on traditional TTS backbones, resulting in limited naturalness and quality. Recent advancements in text-modal language models \cite{devlin2018bert, brown2020language, raffel2020exploring} have further driven the development of language model (LM)-based TTS systems \cite{wang2023neural, zhang2023speak, le2024voicebox, lyth2024natural}. These systems employ pre-trained audio codec models \cite{zeghidour2021soundstream, defossez2022high, kumar2024high} to encode speech waveforms into discrete codes, which are then predicted by prompt-based language models. When trained on large-scale datasets, these models can effectively extract semantic information and synthesize speech that closely approximates human speech in terms of expressiveness and voice quality.

In this paper, we propose a novel CSS framework: DiffCSS. 
Inspired by the success of diffusion models in capturing complex data distributions and generating diverse outputs in the field of computer vision \cite{ho2020denoising, dhariwal2021diffusion, nichol2021improved, rombach2022high}, DiffCSS leverages diffusion models to generate diverse and expressive prosody conditioned on the conversational context. Additionally, DiffCSS integrates an LM-based TTS backbone to synthesize high-quality speech.
In this framework, we first employ a pre-trained codec \cite{ju2024naturalspeech} model to extract prosody-related features from reference speech.
Based on these features, we develop a prosody-enhanced ParlerTTS \cite{lyth2024natural}, which is capable of synthesizing expressive speech conditioned on provided prosody embeddings.
To generate diverse and context-appropriate prosody embeddings from multimodal conversational context, we design a diffusion-based context-aware prosody predictor. 
Experimental results demonstrate that our proposed method significantly outperforms deterministic baselines in terms of expressiveness and contextual coherence. 
Furthermore, the prosody distribution generated by DiffCSS aligns more closely with ground truth distribution, highlighting the effectiveness of incorporating diffusion models into CSS.

To summarize, the main contributions of this paper are:

\begin{itemize}
    \item We proposed a novel CSS framework DiffCSS, which leverages diffusion models to enhance prosody diversity
    depending on conversational context
    along with an LM-based TTS backbone to improve speech quality.
    \item To the best of our knowledge, 
    we are the first to apply diffusion models to CSS systems for conversational context modeling.
    Through prosody diffusion, we significantly enhance the prosody diversity of the synthesized speech.
    \item Through comparative experiments, we demonstrate the effectiveness of our proposed approach in generating speech that is both rich in prosody diversity and suitable for the conversational context.
\end{itemize}

\section{Methodology}

The architecture of our proposed model is illustrated in Fig. \ref{fig:overview}. It consists of two primary components: a TTS backbone based on ParlerTTS \cite{lyth2024natural} and a diffusion-based context-aware prosody predictor. 
The TTS backbone synthesizes high-quality speech based on varying prosody inputs, while the prosody predictor generates diverse prosody embeddings conditioned on the conversational context. 
\subsection{Prosody-enhanced ParlerTTS}

Inspired by the success of LM-based TTS models, we develop a prosody-enhanced ParlerTTS\cite{lyth2024natural} as our TTS backbone, which predicts pre-extracted acoustic tokens by decoder-only transformer blocks.
By adopting the delayed pattern introduced in \cite{copet2024simple}, the TTS backbone is capable of generating high-quality speech autoregressively in a short amount of time. Within the backbone, we designed a prosody extractor to learn prosody embeddings in an unsupervised manner from reference speech. 
To extract prosody features disentangled from other information, we employ the pre-trained FACodec\cite{ju2024naturalspeech} to extract frame-level prosody features $\{F_1, F_2, ..., F_n\}$, where $n$ represents the number of audio frames. 
However, the computation cost of the TTS backbone increases linearly with the length of prosody features.
To reduce resource consumption, we introduce a cross-attention layer with learnable query tokens $\{Q_1, Q_2, ..., Q_m \}$ to derive fixed-length prosody embeddings $\{P_1, P_2, ..., P_m \}$, where $m \ll n$.

These prosody embeddings are then combined with pre-extracted speaker embeddings and serve as the keys and values in the cross-attention layer of the TTS backbone, guiding the speech synthesis process.

\subsection{Diffusion-based context-aware prosody predictor}

\begin{figure}
  \centering
      \includegraphics[scale=0.35]{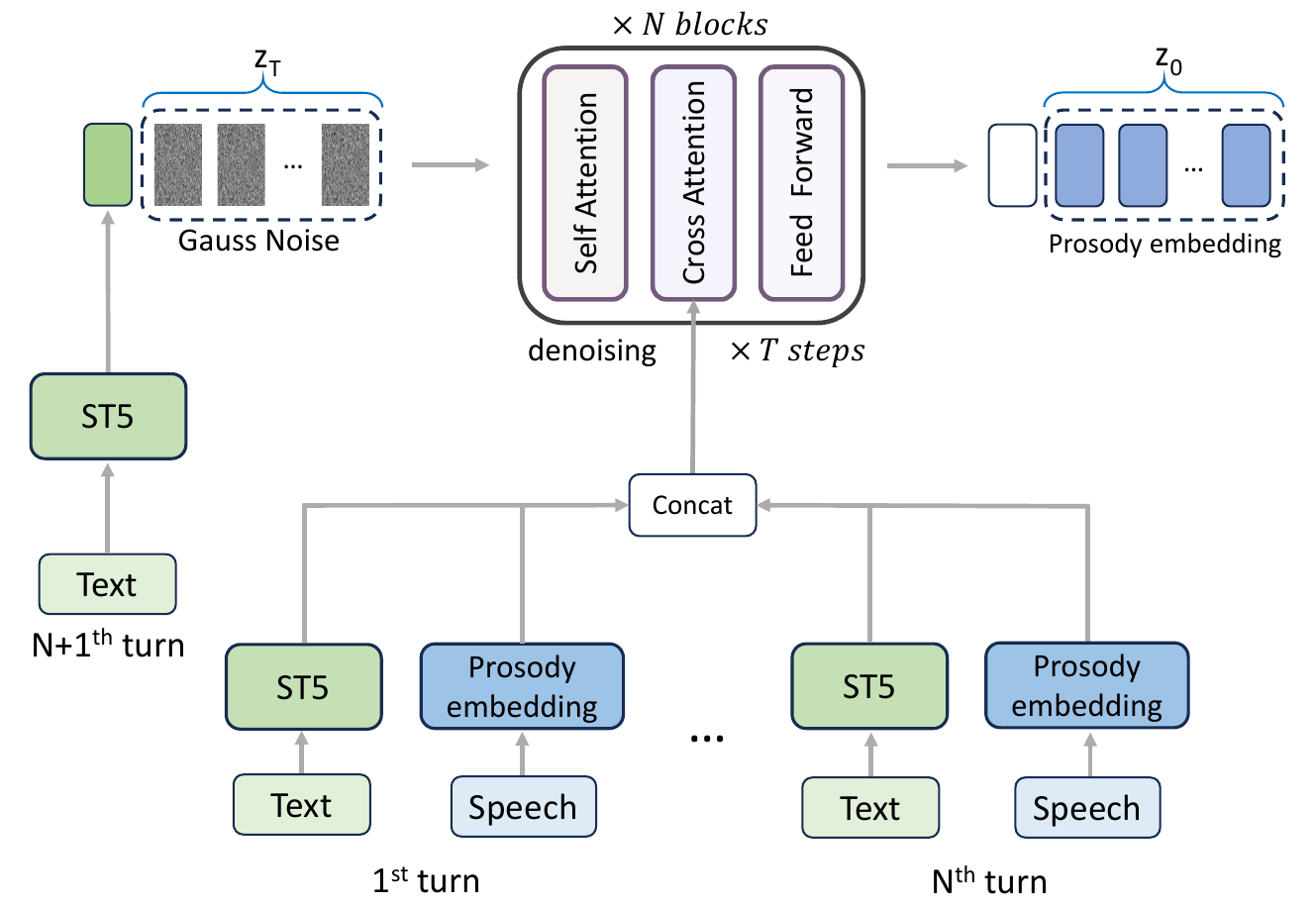}
  \vspace{-6pt}
  \caption{The structure of the diffusion-based context-aware prosody predictor}
  \vspace{-10pt}
  \label{fig:diffusion}
\end{figure}

To predict diverse and contextually appropriate prosody, we design a diffusion-based context-aware prosody predictor that generates the current prosody embedding conditioned on both the multimodal conversational context and the current text. The structure of the prosody predictor is illustrated in Fig. \ref{fig:diffusion}, where a set of Transformer encoder blocks are employed as the underlying denoiser network $\theta$.

For a conversation chunk of length $N + 1$, we combine the multimodal context information as described in Equation \ref{eq:context}, where $c$ denotes the multimodal context information, $s_i$ and $p_i$ denote the textual information extracted by a pre-trained sentence-level T5 \cite{ni2021sentence} and the prosody embedding for the $i$-th turn, respectively. 
\begin{equation}
    c = [s_1, p_1, ..., s_N, p_N]
    \label{eq:context}
\end{equation}

During the diffusion process, Gaussian noise is added to the current prosody embedding $p_{N+1}$ according to a fixed noise schedule $\alpha_1$, ..., $\alpha_T$, where $T$ is the total diffusion steps. This process can be described as Equation \ref{eq:diffusion_1} and \ref{eq:diffusion}, where $\epsilon \sim N(0, I)$ is the Gaussian noise, $z_0$ represents the ground truth prosody embedding $p_{N+1}$ and $\overline{\alpha_t} = \prod \limits_{i=1}^n \alpha_i$.
\begin{gather}
    q(z_t \mid z_{t-1}) = \mathcal{N}\left(z_t; \sqrt{\alpha_t} z_{t-1}, (1 - \alpha_t)I\right) \label{eq:diffusion_1}
    \\
    z_t = \sqrt{\overline{\alpha_t}} z_0 + \sqrt{1 - \overline{\alpha_t}} \epsilon
    \label{eq:diffusion}
\end{gather}


During training, we first uniformly sample a time step $t$ from [1, T], based on which the ground truth prosody embedding $z_0$ is diffused into $z_t$ according to Formula \ref{eq:diffusion}. Then we concatenate textual information of the current sentence $s_{N+1}$ with $z_t$ as the input of the diffusion denoiser. Multimodal context information $c$ is incorporated through cross-attention. Given $z_t$, $s_{N+1}$, $t$ and $c$, the denoiser predicts the added Gaussian noise $\epsilon_\theta(z_t, s_{N+1}, t, c)$ and optimizes its parameters with the following denoising objective:
\begin{align}
    L(\theta) &= E_{t, z_t, \epsilon} \| \epsilon - \epsilon_\theta(z_t, s_{N+1}, t, c) \|_2
\end{align}

In the inference stage, we sample $z_T \sim N(0, I)$ and iteratively perform the denoising process according to Equation \ref{eq:denoising} and \ref{eq:sigma} to compute $z_{t-1}$ from $z_t$ for $t = T, T-1, ..., 1$ where $x \sim N(0, I)$ except for $x=0$ when $t = 1$. After completing T iterations, we obtain the generated prosody embedding $\hat{z_0}$.
\begin{gather}
    z_{t-1} = \frac{1}{\sqrt{a_t}} \left( x_t - \frac{1 - \alpha_t}{\sqrt{1 - \overline{\alpha_t}}} \epsilon_\theta(z_t, s_{N+1}, t, c) \right) + \sigma_t x
    \label{eq:denoising} \\
    \sigma_t = \frac{1 - \overline{\alpha_{t-1}}}{1 - \overline{\alpha_t}} (1 - \alpha_t)
    \label{eq:sigma}
\end{gather}

\subsection{Training Strategy and Inference Procedure}

To enhance the controllability of the TTS module and improve the prosody diversity of synthesized speech, we propose a two-stage training strategy, where prosody is explicitly modeled as an intermediate representation. In order to enhance the semantic understanding and speech synthesis capability of the TTS backbone, we first pre-train the TTS backbone on a large-scale dataset and subsequently fine-tune it on a conversational dataset.
After completing the TTS backbone training, we freeze its parameters and use it to extract ground truth prosody embeddings from reference speech. We then divide the conversations into equal-length chunks, which are used to train the diffusion-based context-aware prosody predictor.

During inference, we first extract the multimodal conversational context information $c$ and the current textual information $s_{N+1}$. We then iteratively compute the predicted prosody embedding $\hat{z_0}$ from Gaussian noise, following Equation \ref{eq:denoising} and \ref{eq:sigma}. Finally, we feed $\hat{z_0}$ into the TTS backbone and synthesize speech according to it.


\section{Experiments}

\subsection{Training setup}
We conduct our experiments on two English datasets. For TTS backbone pre-training, we use an open-source multi-speaker English speech dataset LibriTTS-R \cite{koizumi2023libritts}, which contains 585 hours of high-quality speech from 2,456 speakers. For the conversational dataset, we select DailyTalk \cite{lee2023dailytalk}, which includes 2,541 conversations spanning about 20 hours, performed by two speakers. These conversations are divided into equal-length chunks, each containing 5 utterances, with the first 4 serving as the conversational context for the fifth. The first 2400 conversations are used for training while the remaining 141 are reserved for testing. A pre-trained audio codec model DAC \cite{kumar2024high} is employed to encode the raw waveform with 24kHz sampling rate and reconstruct the waveform based on the predicted acoustic tokens. Speaker embeddings are extracted by a pre-trained voiceprint model\footnote{Available at: \url{https://github.com/modelscope/3D-Speaker/tree/main/egs/3dspeaker/sv-cam++}}.

In our implementation, the TTS backbone comprises 12 transformer decoder blocks, and the diffusion-based context-aware prosody predictor consists of 6 transformer encoder blocks. We pre-train the TTS backbone on LibriTTS-R for 150000 iterations and finetune it on DailyTalk for 20000 iterations with a batch size of 64. The diffusion-based context-aware prosody predictor is trained on DailyTalk for 100000 iterations with a batch size of 32. 

\begin{table*}[ht]
    \vspace{-10pt}
    \caption{The objective and subjective comparisons for different models.}
    \vspace{-0.15cm}
    \centering
    
    \begin{tabular}{lcccccc}
        \toprule
        \textbf{Context modeling method} & \textbf{MOS(Expressiveness)} $\uparrow$ &  \textbf{MOS(Coherence)} $\uparrow$ & \textbf{MCD} $\downarrow$& \textbf{NDB} $\downarrow$ & \textbf{JSD} $\downarrow$ \\
        \midrule
        \textbf{GRU-based} \cite{guo2021conversational} & $3.209 \pm 0.108$ & $3.177 \pm 0.087$ &  8.011 & 16 & 0.227 \\
        \textbf{DialogueGCN-based} \cite{li2022enhancing} & $3.347 \pm 0.104$ & $3.362 \pm 0.093$ & 7.867 & 13 & 0.156 \\
        \textbf{Transformer Encoder-based} & $3.264 \pm 0.112$ & $3.253 \pm 0.103$ & 7.892 & 14 & 0.181 \\
        \textbf{Proposed} & $\textbf{3.602} \pm \textbf{0.101}$ & $\textbf{3.574} \pm \textbf{0.096}$ & \textbf{7.745} & \textbf{4} & \textbf{0.036} \\
        \bottomrule
    \end{tabular}
    \vspace{-15pt}
    \label{tab:main}
\end{table*}

\subsection{Baseline Models}
We implemented the following three models as baselines.

\textbf{GRU-based context modeling} \cite{guo2021conversational} This model uses a unidirectional GRU to process conversational context sequentially.

\textbf{DialogueGCN-based context modeling} \cite{li2022enhancing} This model employs a relation-aware graph convolutional network to model context from both text and audio modalities.

\textbf{Transformer encoder-based context modeling} We implement a contextual prosody predictor based on the Transformer encoder, which shares the same structure and parameters as our proposed method but without diffusion modeling.

\subsection{Subjective Evaluation}

To evaluate the performance of our proposed model in comparison to baselines, we conduct two separate mean opinion score (MOS) tests: one for speech expressiveness and the other for contextual coherence. We randomly select 15 samples and their corresponding contexts from the test set for evaluation. A total of 20 listeners are invited to rate the synthesized speech on a scale of 1 to 5 with 1 point interval, based on both expressiveness and contextual coherence. 

As shown in Table \ref{tab:main}, our proposed method achieves the best E-MOS of 3.602 and C-MOS of 3.574 compared to baselines. This indicates that our model can generate prosody that is both contextually appropriate and expressive. While the Transformer encoder-based method outperforms the GRU-based method, it lags behind the DialogueGCN-based method and significantly underperforms when compared to the proposed method. This suggests that although the Transformer encoder exhibits a moderate ability in modeling conversational context, the integration of diffusion models is essential for improving both prosody prediction and contextual comprehension.

\subsection{Objective Evaluation}

For objective evaluation, we employ Mel-Cepstral Distortion (MCD) to assess the overall quality of the synthesized speech, and Number of Statistically-Different Bins (NDB) along with Jensen-Shannon Divergence (JSD) \cite{richardson2018gans} to evaluate the diversity of prosody, following \cite{huang2023prosody}. NDB and JSD are computed through a clustering process. The procedure is as follows: 1) Cluster the ground-truth prosody into n bins to obtain the ground-truth prosody distribution across bins. 2) Generate prosody samples using the prosody predictor, and assign each generated sample to the closest bin. 3) Calculate the proportion of generated samples in each bin, resulting in a new distribution across the bins. 4) Evaluate the similarity between the generated and ground-truth distributions. JSD is the Jensen-Shannon divergence between the two distributions, and NDB counts the number of bins with statistically significant differences in sample proportions. In this evaluation, we set the number of prosody clustering bins to 20.

As presented in Table \ref{tab:main}, our method achieves an NDB of 4 and a JSD of 0.036, significantly outperforming all baselines. This demonstrates that the generated prosody from our method aligns much more closely with the ground truth prosody distribution. Furthermore, our proposed method also surpasses all baselines in MCD, indicating its ability to synthesize high-quality speech.

\begin{figure}
  \centering
  \includegraphics[scale=0.32]{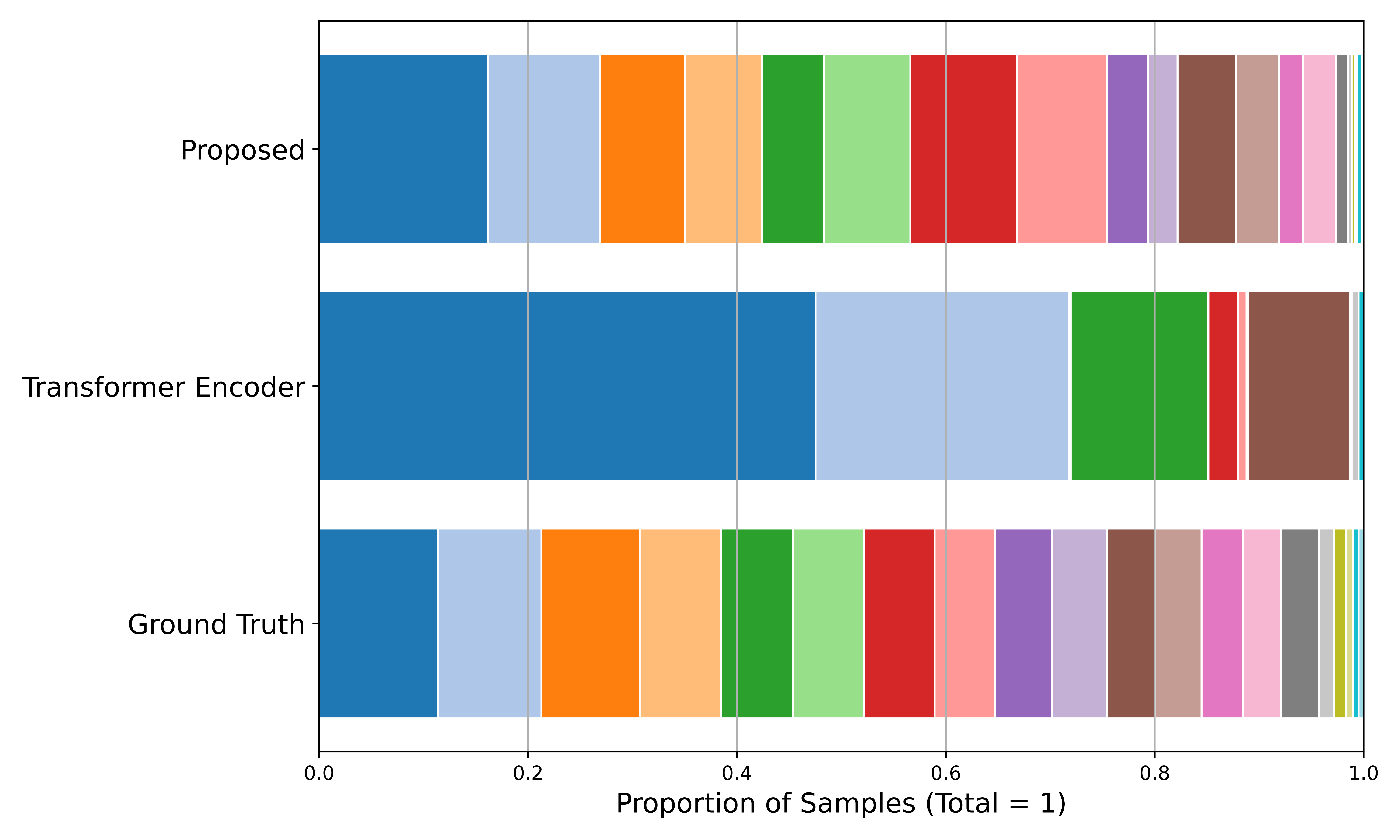}
  \vspace{-10pt}
  \caption{Distribution of prosody embeddings in predictions and ground truth. Each unique color corresponds to a specific cluster.}
  \vspace{-12pt}
  \label{fig:ndb}
\end{figure}

We further visualized the prosody distributions of different methods, as shown in Figure \ref{fig:ndb}, where each unique color corresponds to a specific cluster. The Transformer encoder-based method predicts prosody embeddings concentrated in only a few clusters, with sparse representation in others. In contrast, prosody embeddings generated by our proposed method are more evenly distributed across clusters, closely resembling the ground truth distribution. This indicates that the deterministic baseline tends to predict similar, less diverse prosody, while our proposed method exhibits significantly improved prosody diversity. These results further highlight the importance of incorporating diffusion models in enhancing prosody diversity.

\subsection{Ablation Study on multimodal context}

Furthermore, we investigate the impact of multimodal context by evaluating three different settings: (1) \textbf{Proposed model w/o textual context} (2) \textbf{Proposed model w/o acoustic context} (3) \textbf{Proposed model without full context}. The modalities are excluded by setting the corresponding contextual information to zeros. We conducted a separate training session for each ablation setting, effectively preventing the model from receiving unwanted contextual information.

As shown in Table \ref{tab:ablation}, the absence of either modality reduces the overall quality and prosody diversity of the synthesized speech, thereby diminishing the effectiveness of context modeling. Notably, the exclusion of the acoustic context leads to a greater decline in performance compared to the textual context, underscoring the crucial role of acoustic information in modeling conversational context.

\begin{table}[ht]
    \vspace{-10pt}
    \caption{Ablation studies on context modalities}
    \vspace{-5pt}
    \centering
    \begin{tabular}{lcccccc}
        \toprule
        & \textbf{MCD} $\downarrow$& \textbf{NDB} $\downarrow$ & \textbf{JSD} $\downarrow$ \\
        \midrule
        Proposed & \textbf{7.745} & \textbf{4} & \textbf{0.036} \\
        Proposed w/o textual context & 7.791 & 5 & 0.041 \\
        Proposed w/o acoustic context & 7.946 & 10 & 0.129 \\
        Proposed w/o full context & 8.038 & 12 & 0.152 \\
        \bottomrule
    \end{tabular}
    \label{tab:ablation}
    \vspace{-10pt}
\end{table}

\section{Conclusion}

In this paper, we introduce DiffCSS, a novel CSS framework designed to generate diverse and high-quality speech. We propose a diffusion-based context-aware prosody predictor to generate diverse prosody embeddings conditioned on the conversational context. Additionally, we develop an LM-based TTS backbone to synthesize high-quality speech based on sampled prosody embeddings. Experimental results demonstrate that our proposed model outperforms existing baselines in terms of expressiveness, contextual coherence, and prosody diversity.

\vfill\pagebreak

\bibliographystyle{IEEEbib}
\bibliography{refs}

\end{document}